%% file: Zurita_LHCP_2017.tex
\documentclass[10pt]{article}
\usepackage{graphicx}

\def\Title#1{\begin{center} {\Large #1 } \end{center}}
\def\Author#1{\begin{center}{ \sc #1} \end{center}}
\def\Address#1{\begin{center}{ \it #1} \end{center}}

\newcommand\pubblock{\rightline{\begin{tabular}{l} Proceedings of the Fifth Annual LHCP\\ \pubnumber\\
         \pubdate  \end{tabular}}}

\newenvironment{Abstract}{\begin{quotation} \begin{center} 
             \large ABSTRACT \end{center}\bigskip 
      \begin{center}\begin{large}}{\end{large}\end{center} \end{quotation}}

\newenvironment{Presented}{\begin{quotation} \begin{center} 
             PRESENTED AT\end{center}\bigskip 
      \begin{center}\begin{large}}{\end{large}\end{center} \end{quotation}}

\input econfmacros.tex

\usepackage[usenames, dvipsnames]{xcolor}
\usepackage{multirow}

\newcommand\pubnumber{ TTP17-032 }

\newcommand\pubdate{\today}

\def\affiliation{
Institute for Nuclear Physics (IKP), Karlsruhe Institute of Technology, Hermann-von-Helmholtz-Platz 1, D-76344 Eggenstein-Leopoldshafen, Germany and \\
Institute for Theoretical Particle Physics (TTP), Karlsruhe Institute of Technology, Engesserstra{\ss}e 7, D-76128 Karlsruhe, Germany }

\begin{document}

\large
\begin{titlepage}
\pubblock

\vfill
\Title{  Di-Higgs production at the LHC and beyond  }
\vfill

\Author{ Jos\'e Francisco Zurita  }
\Address{\affiliation}
\vfill
\begin{Abstract}

In this talk I discuss the status and future prospects of testing the Higgs self-couplings at the High Luminosity LHC (HL-LHC) as well as several Beyond Standard Model (BSM) scenarios that could be probed via Higgs pair production in the coming years. 
\end{Abstract}
\vfill

\begin{Presented}
The Fifth Annual Conference\\
 on Large Hadron Collider Physics \\
Shanghai Jiao Tong University, Shanghai, China\\ 
May 15-20, 2017
\end{Presented}
\vfill
\end{titlepage}
\def\thefootnote{\fnsymbol{footnote}}
\setcounter{footnote}{0}
%

\normalsize 


\def\be{\begin{equation}}
\def\ee{\end{equation}}

\section{Introduction}

Di-Higgs production 
is the only LHC process in which the coupling between three Higgs bosons $h$ can be tested at \emph{tree level}. After electroweak symmetry breaking the Higgs potential can be written as
\be
V(h) = \frac{1}{2}m_h^2 h^2 + \lambda v h^3 + \frac{1}{4} \tilde{\lambda} \, .
\ee
In the Standard Model (SM), both $\lambda$ and $\tilde{\lambda}$ are \emph{predicted} once the Higgs mass is known, $\lambda= \tilde{\lambda} \approx 0.13$. Hence \emph{any} deviation automatically implies the existence of Beyond the Standard Model (BSM) phenomena. In addition the knowledge of $\lambda, \tilde{\lambda}$ is crucial to determine the vacuum stability of the Universe~\cite{Degrassi:2012ry}.

In view of the situation above, it is important to understand how accurately the High-Luminosity LHC (HL-LHC) can test these couplings. In the SM, the Higgs pair (triple) production has a cross section of about 40 (0.06) fb~\cite{Borowka:2016ypz} (\cite{Maltoni:2014eza}). Once the corresponding branching fractions are taken into account, the measurement of $\tilde{\lambda}$ needs to wait for a future collider
(see~\cite{Contino:2016spe} and references therein).

I will start by reviewing the status of di-Higgs production in the SM in section~\ref{sec:hhsm}, and then move to a generic discussion of BSM effects in section~\ref{sec:hhbsm}. Finally, I will briefly review a set of studies based on the effect of the Higgs self-coupling in radiative corrections, i.e. \emph{indirect} effects of $\lambda, \tilde{\lambda}$. \footnote{As the activity in the last 5 years on di-Higgs production includes about 200 papers, space limitations forbid a comprehensive overview. Thus I will focus on a few selected examples to illustrate the landscape of New Physics effects in Higgs pair production. }



%

\section{Di-Higgs in the Standard Model}
\label{sec:hhsm}
In the SM the main production mode is gluon fusion, and thus I will focus solely on this channel (VBF and $t\bar{t}hh$ can also be important).
In 1998 this process was known up to next-to-leading order (NLO) in QCD~\cite{Dawson:1998py} in the $m_t \to \infty$ limit. An intense effort in the QCD community lead us to the exact NLO calculation~\cite{Borowka:2016ypz}.

Due to the smallness of the cross section, one must have at least one Higgs decaying into $b \bar{b}$. Since the Higgs discovery several channels have been explored, and the most promising ones have the second Higgs decaying into either $\tau \tau$~\cite{Dolan:2012rv}, $W^+ W^-$~\cite{Papaefstathiou:2012qe}, $\gamma \gamma$\cite{Baglio:2012np}, and $b \bar{b}$~\cite{Behr:2015oqq,deLima:2014dta}. As these studies often used different assumptions, I collect them in table~\ref{tab:smchannels}, where I normalized the result to the value of~\cite{Borowka:2016ypz} and assumed a b-tagging efficiencies of 70 \%.
 ~\footnote{For a real-life exploration of these channels, please see the talks by Harald Fox and David Morse in these proceedings. }.
\begin{table}[!htp]
\begin{center}
\begin{tabular}{l|ccc}
\hline
Process                     & S(600/fb) & B(600/fb) & Reference \\ \hline
$b \bar{b} \tau^+ \tau^ -$   & 50        & 104     & 
{\rm \tiny \textcolor{Blue}{Dolan, Englert, Spannowsky}~\cite{Dolan:2012rv}} \\  \hline
$b \bar{b} W^+ W^-$          & 12        & 8      & 
{\rm \tiny \textcolor{Blue}{Papaefstathiou, L.L. Yang, Zurita}~\cite{Papaefstathiou:2012qe}} \\   \hline
\multirow{2}{*}{$b \bar{b} \gamma \gamma$ } & 9         & 11      & {\rm \tiny \textcolor{Blue}{Baglio, Djouadi, Gr{\"o}ber, M\"uhlleitner, Quevillon, Spira}~\cite{Baglio:2012np}}   \\ \cline{2-3}
                        & 6         & 12.5    & 
{\rm \tiny \textcolor{Blue}{Baur, Plehn, Rainwater}~\cite{Baur:2003gp}}   \\ \hline
\multirow{2}{*}{$b \bar{b} b \bar{b}$}   & 48        & 2000  & {\rm \tiny \textcolor{Blue}{Behr, Bortoletto, Frost, Hartland, Issever, Rojo}~\cite{Behr:2015oqq}}     \\ \cline{2-3}
                        & 50        & 2500 & {\tiny \tiny \textcolor{Blue}{Ferreira da Lima, Papaefstathiou, Spannowsky}~\cite{deLima:2014dta}} \\  \hline
                        
                        \end{tabular}
\caption{Pheno studies of the di-Higgs process in several final states. We normalized all the studies to the cross section of Ref.~\cite{Borowka:2016ypz} and assumed a b-tagging efficiency of 70 \% with a 1\% light jet rejection. }
\label{tab:smchannels}
\end{center}
\end{table}
\vspace{-0.5cm}

From the table we see that we have at our disposal only a handful of signal and background events, as anticipated. Hence the sensitivity in a full-fledged analysis can drastically vary with respect to these crude estimations. However, it is hard to avoid being optimistic when considering a combination of all these final states at the HL-LHC.



\section{Beyond Standard Model effects in di-Higgs production}
\label{sec:hhbsm}
\begin{figure}[htb]
\centering
\includegraphics[height=1.2in,width=0.8\textwidth]{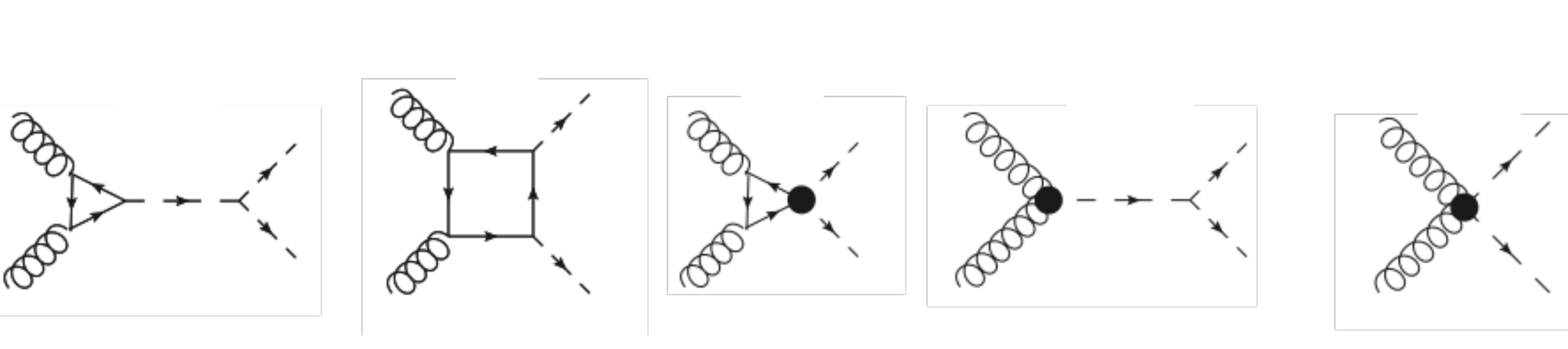}
\caption{$g g \to h h$ production in BSM theories. }
\label{fig:ggtohh}
\end{figure}

We start by depicting the production diagrams for di-Higgs in a generic BSM theory, in figure~\ref{fig:ggtohh}. 

The main effects in di-Higgs production arise from
:
\begin{itemize}
\item Anomalous couplings.
\item New states running in the loop.
\item New resonances.
\item Chain decays, $Y \to h X$.
\item Exotic Higgs decays.
\end{itemize}

\subsubsection*{Anomalous couplings}
\begin{figure}[htb]
\centering
\includegraphics[height=1.5in,width=0.35\textwidth]{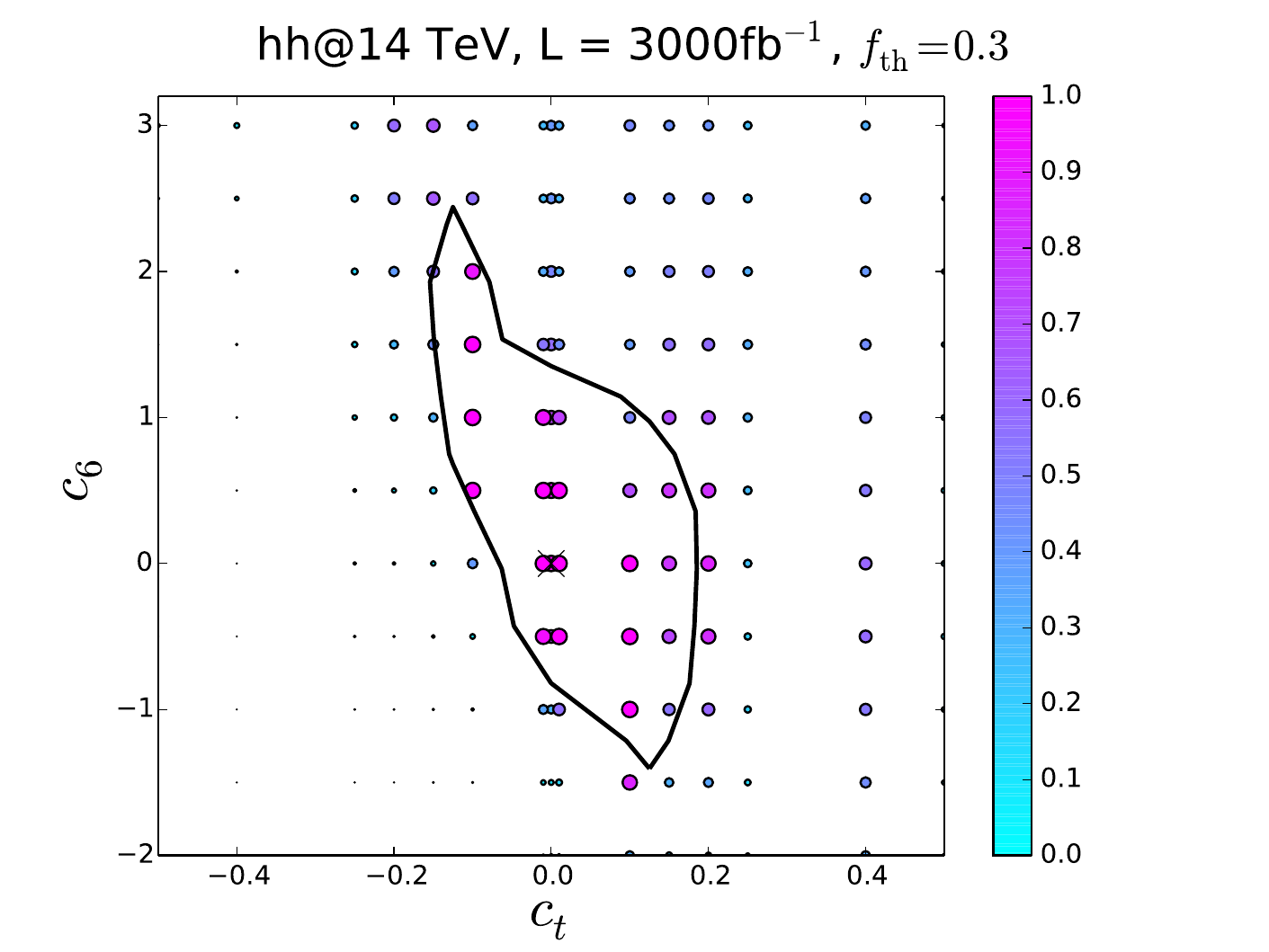}
\includegraphics[height=1.5in,width=0.30\textwidth]{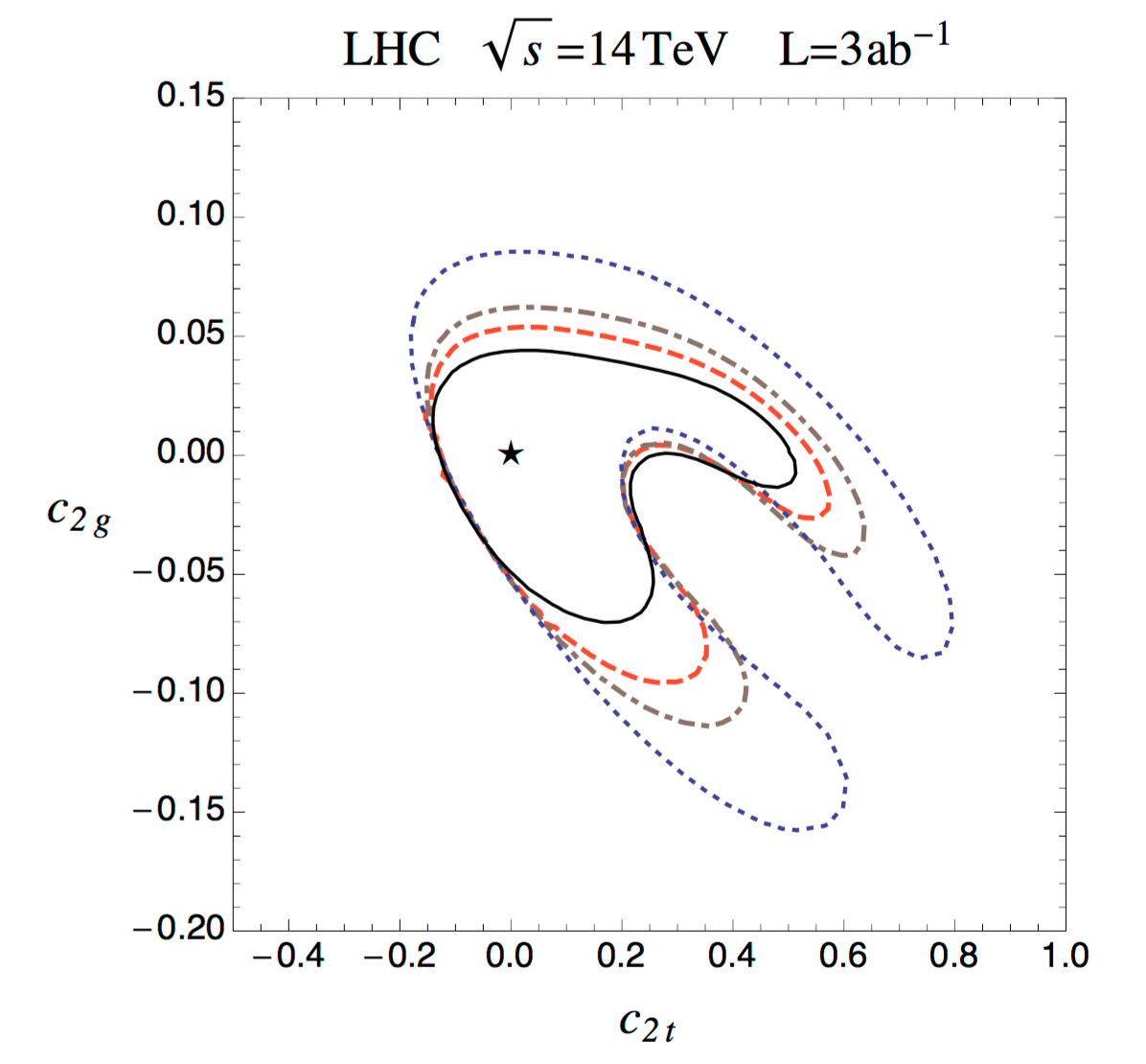}
\caption{ 2-D parameter spaces constrained with the HL-LHC for $c_6-c_t$ (left,~\cite{Goertz:2014qta}) and $c_{2g}-c_{2t}$~(right,\cite{Azatov:2015oxa}). }
\label{fig:anomEFT}
\end{figure}

Not only the $hhh~(\kappa_{h})$ coupling matters, but also $ht\bar{t}~(\kappa_t), hht\bar{t}~(\kappa_{2t}), gghh~(\kappa_{2g})$, ... (as seen in figure~\ref{fig:ggtohh}). 
These can be considered free parameters or can be correlated if they arise from a concrete Lagrangian, either a complete model or in an effective field theory (EFT) construction, in which case the $\kappa_i$ coefficients become simple functions of the $c_i$ Wilson coefficients. We show in figure~\ref{fig:anomEFT} the HL-LHC reach for two 2-D $c_i$ planes (from~\cite{Goertz:2014qta} and \cite{Azatov:2015oxa}). These plots illustrate that the different coefficients can be highly correlated. In particular, the left panel shows a correlation between $c_6$ and $c_t$~\footnote{The notation for the Wilson coefficients follows references~\cite{Goertz:2014qta} and \cite{Azatov:2015oxa}.}, thus casting doubts on the \emph{$c_6$-only} approach that most indirect studies (see sec.~\ref{sec:indi}) follow to present their results.

\subsubsection*{Colored states running in the loop}
The triangle and box diagrams scale differently from the SM case when the loop features new colored particles. These are ubiquitous in several BSM scenarios, like Composite Higgs or Supersymmetry (see e.g: \cite{Kribs:2012kz,Dawson:2015oha,Batell:2015koa}). 

\begin{figure}[htb]
\centering
\includegraphics[height=1.6in,width=0.45\textwidth]{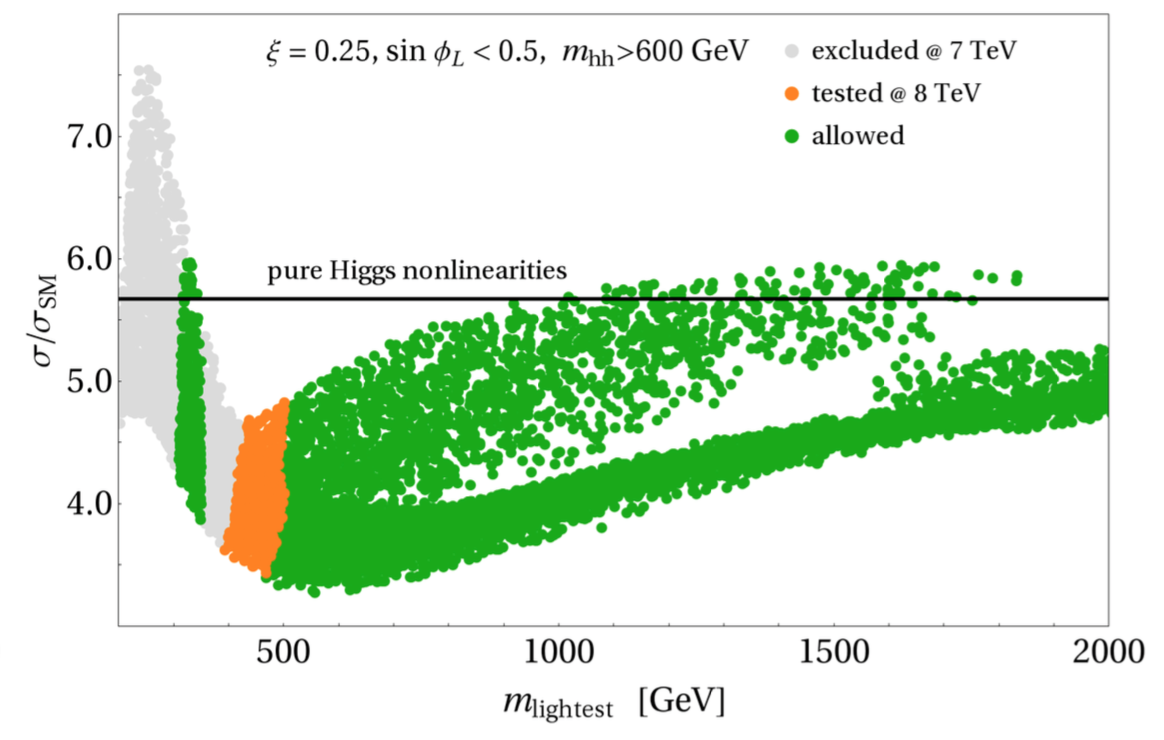}
\includegraphics[height=1.6in,width=0.45\textwidth]{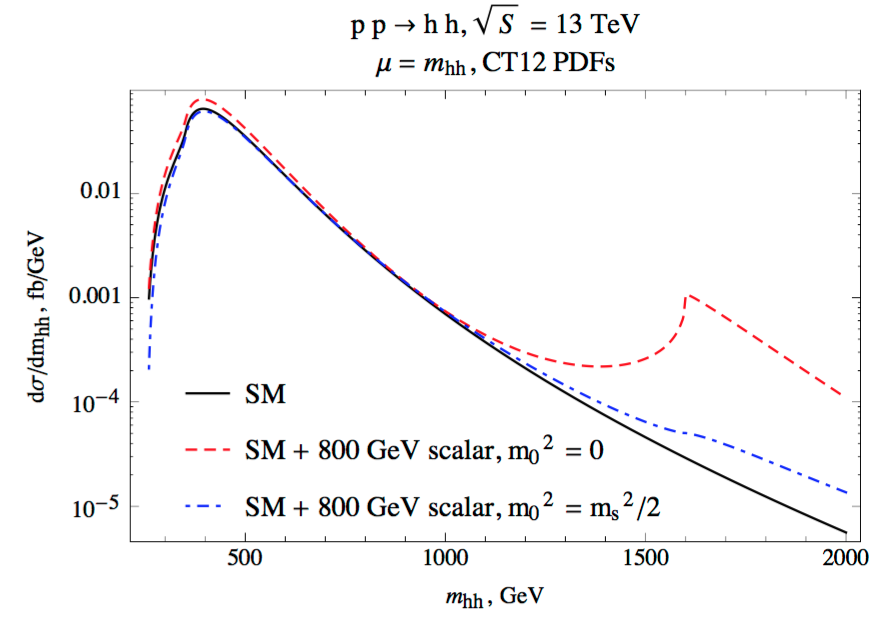}
\caption{Effects of new colored states in (left) inclusive cross section~\cite{Gillioz:2012se}) and (right) $m_{hh}$ distribution~\cite{Dawson:2015oha}. }
\label{fig:loops} 
\end{figure}

The presence of these new states affects both inclusive cross sections and differential distributions, as shown in figure~\ref{fig:loops}. In the left panel we see, in the context of a concrete Composite Higgs model, that the cross section can be 4-5 times larger than the SM one, even for fairly heavy (2 TeV) new particles. For vector -like quarks the $m_{hh}$ distribution is practically unaffected~\cite{Dawson:2015oha}, while for scalars and chiral fermions one finds large deviations in the tails, as illustrated in the right panel of fig.~\ref{fig:loops}. 
\subsubsection*{New resonances}
\begin{figure}[htb]
\centering
\includegraphics[height=1.6in,width=0.35\textwidth]{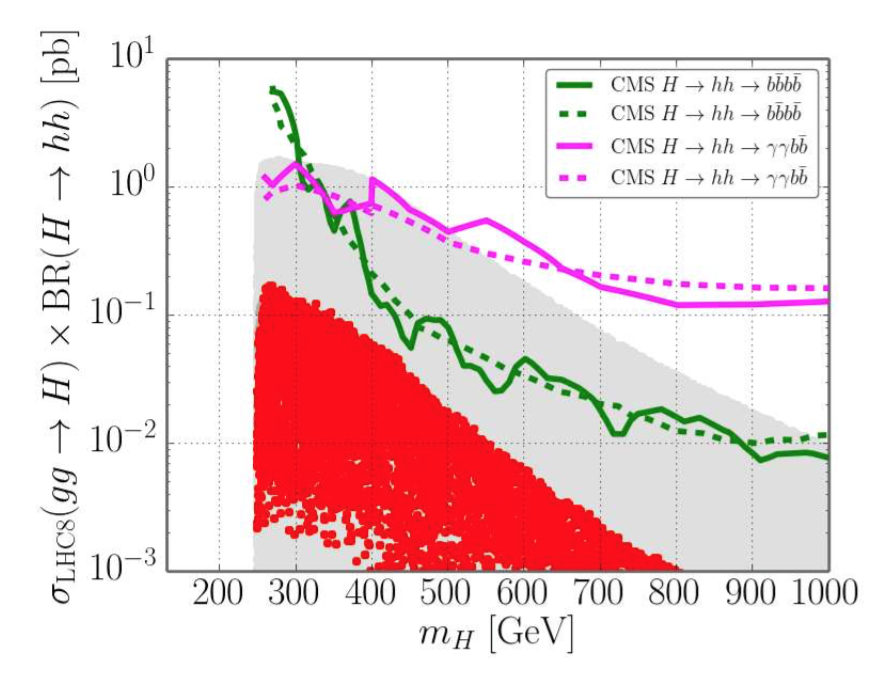}
\includegraphics[height=1.6in,width=0.35\textwidth]{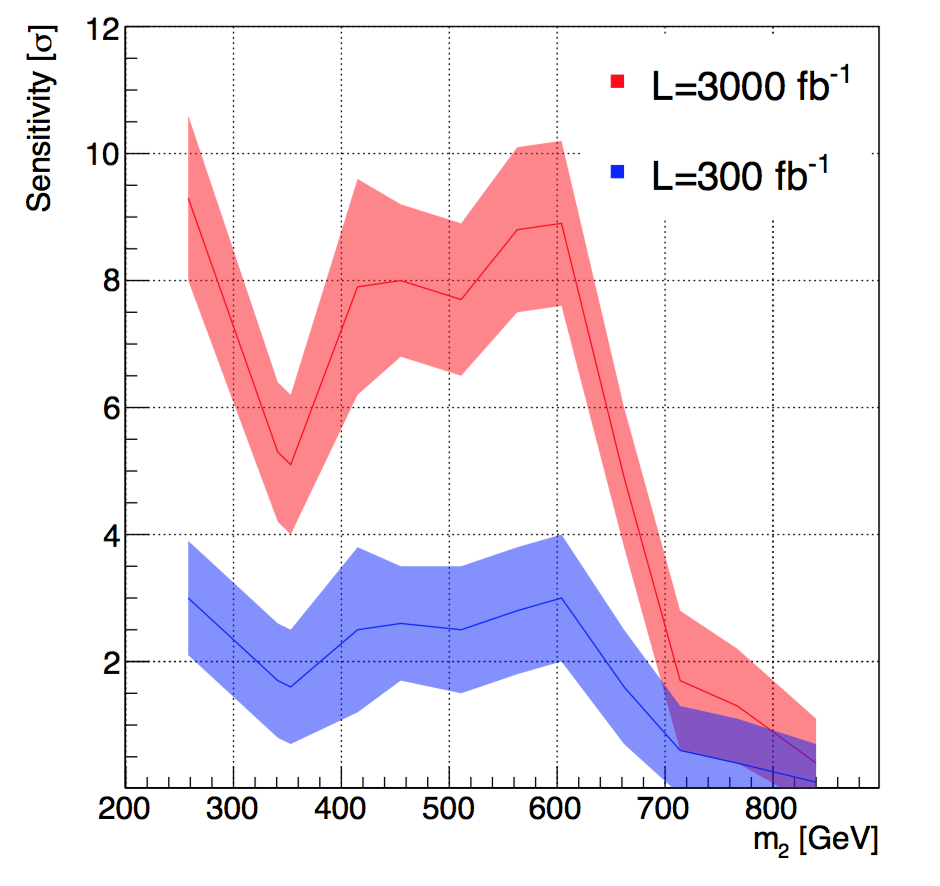}
\caption{Left: Total rate for $S \to hh$ including the bounds from $hh$ resonant searches~\cite{Robens:2016xkb}). Right: Significance vs $m_H$ in a model allowing first order electroweak phase transition~\cite{Huang:2017jws}.}
\label{fig:singlet}
\end{figure}

A new scalar resonance S coupling to two Higgs bosons appear in the context of extended Higgs sectors, like in Two Higgs doublet models~\cite{Dolan:2012ac} or in  Higgs portal models~\cite{No:2013wsa}. In the latter (the simplest scalar extension of the SM) $S$ and the SM Higgs mix with an angle $\alpha$ into the $h(125)$ and $H(m_H)$ states.
The free parameters of this model are the mixing angle $\alpha$, the heavy Higgs mass $m_H$ and the ratio of the SM Higgs and S vacuum expectation values. We show in the left panel of fig~\ref{fig:singlet} how the currently allowed region will be probed by $H \to hh$ searches in the near future. In the right panel we show a subset of the parameter space, namely the case where the extra singlet is used to obtain a first order electroweak phase transition, thus leaving only one free parameter $m_2=m_H$. Using only the fully leptonic $b\bar{b}W^+W^-$ final state one obtains 95 \% C.L exclusions up to 700 GeV using the HL-LHC, thus providing a robust test of scenarios featuring electroweak baryogenesis.
\vspace{-0.5cm}
\subsubsection*{Chain decays and exotic final states}
Higgs chain decays are tightly constrained by the $Y \to Z X$ process (see however~\cite{Kang:2015nga}). Exotic Higgs decays are complicated due to the existing (and expected) bounds, but this was attempted in ref.~\cite{Banerjee:2016nzb}.

\section{Going beyond HH: indirect effects}
\label{sec:indi}
While $\lambda$ enters at tree level in $p p \to h h$, one could look for the effect of $\lambda$ in loops. In the recent years this possibility was raised by McCullough~\cite{McCullough:2013rea} in the $e^+ e^- \to H Z$ process. During the last year, several studies considered the effects of radiative corrections involving $\lambda$ in several observables, which we summarize in table~\ref{tab:table2}. These works assume that only one operator, namely $\frac{c_6}{\lambda_{\rm SM}}{v^2} \big(H^{\dagger} H) ^ 3$ is added to the SM.
\begin{table}[!htbp]
\begin{center}
\begin{tabular}{l|ccc}
\hline
Observable(s)                                                & $c_6$  @ $95\% $C.L   & Reference                                        \\ \hline
gg $\to$ hh @ 13 TeV                                        & {[}-9.4,12.4{]}  & { \tiny \textcolor{Blue}{ATLAS-CONF-2016-049}~\cite{ATLAS:2016ixk}}                              \\ \hline
gg $\to$ h, h $\to \gamma \gamma$                                  & {[}-12.7,9.9{]}  & { \tiny \textcolor{Blue}{Gorbahn, Haisch 1607.03773}~\cite{Gorbahn:2016uoy}  }                       \\ \hline
gg $\to$ h, VBF ; h $\to \gamma \gamma, ZZ, WW, \tau \tau$ & {[}-10.4,16.0{]} & { \tiny \textcolor{Blue}{de Grassi, Giardino, Maltoni, Pagani 1607.04251} ~\cite{Degrassi:2016wml}} \\ \hline
Vh, VBF                                                   & {[}-13.0,15.3{]} & { \tiny \textcolor{Blue}{Bizon, Gorbahn, Haisch, Zanderighi 1610.05771}~\cite{Bizon:2016wgr}}    \\ \hline
S,T parameters                                            & {[}-15.0,16.4{]} & { \tiny \textcolor{Blue}{Kribs, Maier, Rzehak,Spannowsky, Waite 1702.07678}~\cite{Kribs:2017znd}} \\ \hline
$m_W, s_\theta^2$                                         & {[}-15.0,16.0{]} & { \tiny \textcolor{Blue}{de Grassi, Fedele, Giardino 1702.01737} ~\cite{Degrassi:2017ucl}       }   \\ \hline
\end{tabular}
\caption{Constraints at the 95 \% C.L in the Wilson coefficient $c_6$ from different indirect effects. Note that all these results have been derived under the assumption that only }
\label{tab:table2}
\end{center}
\end{table}
Note that while these results seem to suggest that indirect effects are more constraining than the direct search for di-Higgs, this is only true if correlations can be neglected. The final answer would come from an ultimate global fit to all relevant observables, as carried out in~Ref.~\cite{DiVita:2017eyz} yielding $\lambda / \lambda_{\rm SM} \in [-1.8,7.5]$ at the 95 \% C.L.
\section{Conclusions}
In the last years a previoulsy considered \emph{impossible measurement} has received a lot of attention, 
in light of the 125 GeV Higgs boson properties. 
From the BSM angle, there is a very rich phenomenology, compatible with current constraints but still giving large deviations in di-Higgs associated observables. Finally, while the LHC can scratch the surface of the triple Higgs coupling, a full reconstruction of the SM Higgs potential requires $\tilde{\lambda}$ and (most likely) a more accurate determination of $\lambda$, which is only possible at future colliders~\cite{Contino:2016spe}.


\end{document}

%% file: econfmacros.tex
\def\beq{\begin{equation}}
\def\eeq#1{\label{#1}\end{equation}}
\def\eeqn{\end{equation}}

\def\beqa{\begin{eqnarray}}
\def\eeqa#1{\label{#1}\end{eqnarray}}
\def\eeqan{\end{eqnarray}}

\let\bar=\overbar

\def\Dslash{\not{\hbox{\kern-4pt $D$}}}
\def\dslash{\not{\hbox{\kern-2pt $\del$}}}

\def\ee{e^+e^-}

\def\msb{{\bar{\ssstyle M \kern -1pt S}}}

